**Title:**

Skill-Based Labor Market Polarization in the Age of AI: A Comparative Analysis of India and the United States

Venkat Ram Reddy Ganuthula[1]   Krishna Kumar Balaraman[1]

Indian Institute of Technology Jodhpur

**Abstract:**

This paper examines labor market polarization through a comparative analysis of skill-based employment and wage distributions in India and the United States during 2018-2023, with particular attention to differential automation risks and AI preparedness. Using detailed occupation-level data, automation risk metrics, and a series of statistical tests including wage premium analysis, employment share tests, and wage-employment regressions, we document significant structural differences in labor markets between developing and developed economies. Our analysis yields four key findings. First, we find statistically significant differences in employment distribution patterns, with India showing disproportionate concentration in low-skill employment compared to the US, particularly in occupations with high automation risk. Second, regression analysis reveals that wage premiums differ systematically between the two countries, with significantly larger skill-based wage gaps in India. Third, we find robust evidence of a negative relationship between employment size and wages, suggesting stronger labor supply effects in developing economies. Fourth, analysis of occupation-specific automation risk reveals that developing economies face a "double vulnerability" - concentration of employment in both low-skill occupations and jobs with higher automation potential, complicated by lower AI preparedness scores. These findings provide novel empirical evidence on how development stages influence labor market polarization patterns and carry important implications for skill development and technology adoption policies in developing economies. Our results suggest that traditional approaches to labor market development may need significant modification to account for the differential impacts of AI across development stages.

# 1. Introduction

Labor markets worldwide are experiencing unprecedented transformations, driven by both traditional development forces and emerging technological disruptions. While substantial research documents the polarization of labor markets in developed economies (Autor, Katz, and Kearney 2006; Goos, Manning, and Salomons 2014), characterized by declining middle-skill employment and rising wage inequality, our understanding of how these patterns manifest in developing economies remains limited (Fleisher et al., 2018). This gap is particularly significant as developing economies now face a dual challenge: managing traditional development transitions while simultaneously preparing for artificial intelligence (AI)-driven labor market transformation.

The relationship between economic development and labor market structure raises fundamental questions that build on seminal work in development economics (Lewis 1954; Harris and Todaro 1970). Recent research suggests that technological change and globalization may affect labor markets differently across development stages (Goldberg and Pavcnik 2007; Acemoglu and Autor 2011). The emergence of AI and automation technologies adds new complexity to this dynamic. While automation risk varies significantly across occupations (Frey and Osborne 2017), countries differ markedly in their AI preparedness and absorption capabilities (World Bank 2021), suggesting potential divergence rather than convergence in labor market outcomes.

Understanding these patterns is crucial for several reasons. First, it provides insights into whether theories of labor market polarization, primarily developed in the context of advanced economies (Autor and Dorn 2013), apply uniformly across development stages in an era of rapid technological change. Second, it helps identify whether developing economies face unique structural challenges that require different policy approaches, particularly given varying levels of AI readiness (Banerjee and Duflo 2011). Third, it contributes to our understanding of how labor market institutions, development stages, and technological capabilities interact to shape employment and wage outcomes (Freeman 2010).

This paper addresses these issues through a comparative analysis of labor markets in India and the United States, incorporating novel data on occupation-specific automation risk and country-level AI preparedness. Following methodological approaches established in cross-country labor market studies (Ashenfelter and Card 2011; Katz and Autor 1999), we examine how development stages influence labor market structure in the context of emerging technological capabilities. Our focus on India and the US builds on previous comparative work (Hsieh and Klenow 2009) while extending it to consider differential automation risks and technological absorption capacities.

Our analysis focuses on four primary research questions that emerge from current debates in the literature:

1. How do skill-based employment distributions systematically differ between developing and developed economies, and to what extent are these differences concentrated in occupations with varying automation risk? This question extends work by Autor (2015) and Das and Kar (2018) on employment polarization while incorporating automation vulnerability.
2. To what extent do wage premiums across skill levels vary with economic development, and how might these premiums be affected by differential AI adoption capabilities? This builds on research examining skill premiums in developing countries (Card, Kramarz, and Lemieux 1999) while considering technological readiness.
3. Does labor market polarization follow different patterns in developing versus developed economies when accounting for automation potential? This question connects to broader debates about convergence in labor market institutions (Freeman and Oostendorp 2020) and technological adoption patterns.
4. How does the interaction between development stage and AI preparedness affect the vulnerability of different occupational categories? This new question emerges from recent work on automation risk (Acemoglu and Restrepo 2019) and technological absorption in developing economies.

To address these questions, we employ a comprehensive empirical strategy combining distributional analysis, wage premium estimation, and regression analysis, following methodological advances in labor economics (Angrist and Pischke 2009). Our approach allows us to test for systematic differences in employment distributions, wage determination patterns, and labor market polarization between the two countries while accounting for occupation-specific automation risks and country-level AI capabilities.

This study contributes to multiple strands of literature. First, it extends work on labor market polarization (Acemoglu and Autor 2011) by providing systematic evidence on how polarization patterns vary with development stages and technological readiness. Second, it contributes to development economics by documenting how labor market structures evolve with economic development in the context of emerging technologies (Rodrik 2016). Third, it provides insights for policy by identifying specific areas where developing economy labor markets differ systematically from developed ones, particularly in their vulnerability to technological disruption (World Bank 2019). Fourth, it adds to the growing literature on automation and labor markets (Frey and Osborne 2017) by examining how automation risk interacts with development stage to shape labor market outcomes.

The remainder of the paper is organized as follows. Section 2 reviews relevant literature and develops our theoretical framework. Section 3 describes our data and empirical methodology, including our novel datasets on automation risk and AI preparedness. Section 4 presents our empirical analysis and results. Section 5 discusses policy implications, with particular attention to the challenges of simultaneous development and technological adaptation. Section 6 concludes.

## 2. Literature Review

This paper intersects with several important strands of literature in labor economics and development studies. We begin by examining research on labor market polarization in developed economies, which provides the theoretical foundation for understanding changing employment patterns. We then review literature on wage determination in developing countries, which reveals distinct patterns from advanced economies. This is followed by analysis of comparative labor market institutions, which helps explain structural differences across development stages. Finally, we examine emerging research on AI-driven labor market transformation, which suggests new dynamics in skill premiums and employment patterns.

### 2.1 Labor Market Polarization in Developed Economies

The phenomenon of labor market polarization was first systematically documented in the United States labor market through groundbreaking research by Autor et al. (2003), who demonstrated increasing employment shares in both high-skill and low-skill occupations, coupled with declining middle-skill employment. This "hollowing out" of the middle class sparked extensive research across developed economies. Subsequent studies by Goos et al. (2009) and Michaels et al. (2014) confirmed similar patterns in European labor markets, suggesting a broader structural transformation in advanced economies.

The theoretical explanations for these patterns have evolved along several lines. Autor and Dorn (2013) developed the routine-biased technological change (RBTC) framework, arguing that computerization disproportionately affects middle-skill routine tasks. This framework helped explain why technological advancement might simultaneously benefit high-skill cognitive work while preserving low-skill manual tasks. Building on this foundation, Acemoglu and Autor (2011) proposed a more comprehensive task-based framework that explains polarization through the interaction of technological change and international trade, providing a richer understanding of the forces shaping employment distributions.

Recent research has revealed significant heterogeneity in polarization patterns across regions and sectors. Autor et al. (2020) documented substantial variation across U.S. metropolitan areas, finding that local labor market conditions significantly influence polarization dynamics. Similarly, Heyman (2016) identified diverse patterns across European countries, suggesting that institutional frameworks and industrial structures mediate the impact of technological change. These findings highlight the importance of considering local context when analyzing labor market transformation.

The emergence of artificial intelligence has prompted a reevaluation of traditional polarization theories. Recent work by Webb (2020) suggests that AI's impact may differ fundamentally from previous waves of technological change, potentially affecting high-skill cognitive tasks that were previously thought immune to automation. Acemoglu et al. (2022) further develop this perspective, arguing that AI-driven automation may create new patterns of labor market transformation that deviate from historical experience.

**2.2 Wage Determination in Developing Countries**

The analysis of wage determination in developing economies builds on foundational work by Mincer (1974), who established the theoretical framework for understanding returns to education and skill. This seminal contribution has shaped decades of research on human capital and wage premiums. Subsequent studies by Psacharopoulos and Patrinos (2018) have consistently documented higher returns to education in developing countries, though the mechanisms driving these elevated returns remain debated.

The relationship between skills and wages in developing economies is complicated by several structural factors. Goldberg and Pavcnik (2007) emphasize the role of large informal sectors and different institutional frameworks in shaping wage determination patterns. Their research highlights how traditional models of wage determination may need modification when applied to developing country contexts. These insights suggest that the interaction between skills, technology, and wages may follow different patterns in developing versus developed economies.

Recent empirical work has begun to examine how technological change affects wage determination in developing economies. Das and Kar (2018) find that digital technology adoption in India has led to increasing returns to cognitive skills while simultaneously creating new forms of wage inequality. This research suggests that technological change may exacerbate existing labor market disparities in developing economies. Similarly, Vashisht and Dubey (2019) document significant variation in automation potential across sectors in developing economies, with important implications for wage determination patterns.

**2.3 Comparative Labor Market Institutions**

Institutional frameworks play a crucial role in shaping labor market outcomes across different development stages. Freeman and Oostendorp (2020) provide comprehensive documentation of systematic differences in wage-setting institutions between developed and developing economies. Their research reveals how institutional variations persist even as economies develop, suggesting deep structural differences in labor market organization.

The relationship between institutional frameworks and development levels has been extensively studied. Botero et al. (2004) demonstrate that labor market regulation varies systematically with economic development, while Heckman and Pagés (2004) find that these institutional differences significantly affect wage and employment outcomes. These findings suggest that institutional frameworks may create persistent differences in labor market functioning across development stages.

Recent research has begun to examine how these institutional differences interact with technological change. Alvarez et al. (2018) find evidence that developing countries' labor market institutions may amplify rather than mitigate the effects of technological disruption. This amplification effect could have important implications for how developing economies navigate technological transitions. The World Bank's Digital Development Report (2021) further highlights how varying levels of digital infrastructure and regulatory frameworks affect countries' ability to adapt to technological change.

## 2.4 AI and Automation in Labor Markets

### 2.4.1 Differential Impact Across Skill Levels

The impact of artificial intelligence on labor markets appears to deviate from traditional patterns of skill-biased technological change. Webb (2020) provides evidence that AI capabilities may complement some middle-skill occupations while potentially automating certain high-skill cognitive tasks. This finding challenges conventional wisdom about the relationship between skill levels and automation risk, suggesting more complex patterns of technological displacement.

### 2.4.2 Variation in AI Preparedness

The World Economic Forum's Global AI Readiness Index (2023) reveals substantial variation in countries' capabilities to implement AI technologies effectively. This variation extends beyond traditional measures of digital infrastructure to encompass factors such as regulatory frameworks, education systems, and innovation ecosystems. These differences in AI preparedness may have important implications for the distribution of benefits from AI adoption across countries.

### 2.4.3 Development Stage Interaction

Emerging theoretical work by Korinek and Stiglitz (2021) suggests that AI adoption may lead to divergence rather than convergence between developed and developing economies. This divergence could result from differences in absorption capacity and complementary factors of production. Their analysis raises important questions about whether AI technology might exacerbate rather than reduce international inequality.

## 2.5 Synthesis and Research Gaps

The existing literature reveals several important gaps in our understanding of how labor markets evolve across development stages in the context of technological change. While labor market polarization is well-documented in developed economies, we lack systematic evidence on how these patterns manifest in developing economies when accounting for AI and automation potential. Similarly, although studies examine wage determination in developing countries, few directly compare wage-setting patterns across development stages while considering differential automation risks.

Our review also identifies gaps in understanding how institutional differences between developed and developing economies affect AI adoption and labor market outcomes. While institutional variations are well-documented, their implications for technological absorption and labor market transformation remain underexplored. Furthermore, the interaction between development stage, AI preparedness, and labor market vulnerability has not been systematically analyzed in a comparative framework.

This paper addresses these gaps through several contributions. First, we provide a systematic comparison of labor market structures between India and the US while incorporating occupation-specific automation risks. Second, we examine how wage determination patterns vary with development stage and technological readiness. Third, we analyze how institutional differences affect both traditional labor market outcomes and AI adoption patterns. Finally, we develop a framework for understanding how development stage and AI preparedness interact to shape labor market vulnerability.

## 3. Data and Methodology

### 3.1 Data Sources and Sample Construction

Our empirical analysis draws on multiple data sources that combine traditional labor market metrics with novel data on automation risk and AI preparedness. Following the methodological approach of Card et al. (2018), we construct a comprehensive dataset that allows for systematic comparison of labor market outcomes across development stages while accounting for technological factors.

The primary employment and wage data for India comes from the National Sample Survey Office (NSSO) Employment-Unemployment Survey and Periodic Labour Force Survey (PLFS), covering 63 distinct occupations over the period 2018-2023. Following Vashisht and Dubey (2019), we harmonize these occupational categories with international classifications to ensure comparability. For the United States, we utilize the Current Population Survey (CPS) and Occupational Employment Statistics (OES), following the data construction approach of Autor and Dorn (2013).

Our automation risk metrics build on the methodological framework developed by Frey and Osborne (2017), supplemented with task-level data from O*NET (Occupational Information Network). Following Webb (2020), we construct occupation-specific automation risk scores that incorporate both traditional automation potential and AI-specific capabilities. The scores are standardized on a scale of 1-70, with higher values indicating greater automation potential.

The measurement of AI preparedness draws on the World Economic Forum's Global AI Readiness Index (2023), which provides comprehensive country-level assessments across multiple dimensions. Following Korinek and Stiglitz (2021), we focus particularly on measures of digital infrastructure capability, human capital development, and innovation ecosystem maturity. This approach allows us to capture both technological and institutional aspects of AI readiness.

### 3.2 Variable Construction and Measurement

Our employment measures are constructed following the methodology of Autor et al. (2020), with employment shares calculated at both the occupation and skill-level for each country-year observation. Following Acemoglu and Autor (2011), we classify occupations into three skill categories (high, medium, low) based on educational requirements and task complexity. This classification is harmonized across countries using the International Standard Classification of Occupations (ISCO-08) framework.

Wage measures are constructed following the approach of Katz and Murphy (1992), with particular attention to addressing measurement challenges in developing economy contexts as highlighted by Goldberg and Pavcnik (2007). We calculate real wages using country-specific consumer price indices and convert to purchasing power parity (PPP) dollars following the methodology of Freeman and Oostendorp (2020). This ensures comparability across countries while accounting for differences in price levels and living standards.

The automation risk measures combine multiple dimensions of technological vulnerability. Following the approach of Webb (2020), we construct a composite index that incorporates both traditional automation risk metrics and AI-specific capabilities. The index is validated using the methodology of Arntz et al. (2017), which emphasizes task-level analysis rather than occupation-level aggregates.

### 3.3 Empirical Strategy

Our empirical analysis proceeds through several stages, each addressing specific aspects of labor market transformation across development stages. Following Acemoglu and Autor (2011), we begin with the baseline employment distribution specification:

$$\ln(E_{ijkt}) = \alpha + \beta\ C_i + \beta\ S_j + \beta\ A_k + \beta\ (C_i \times S_j) + \beta\ (C_i \times A_k) + \beta\ (S_j \times A_k) + \gamma X_{ijkt} + \varepsilon_{ijkt} \quad (1)$$

where:

- $E_{ijkt}$ represents employment in country i, skill level j, automation risk category k, at time t
- $C_i$ is a country indicator
- $S_j$ represents skill level indicators
- $A_k$ denotes automation risk category
- $X_{ijkt}$ is a vector of control variables
- $\varepsilon_{ijkt}$ is the error term

For wage premium analysis, we employ the following specification, building on Mincer (1974) and extending it to incorporate technological factors:

$$\ln(W_{ijkt}) = \alpha + \beta\ S_j + \beta\ C_i + \beta\ A_k + \beta\ P_i + \beta\ (S_j \times C_i) + \beta\ (S_j \times A_k) + \beta\ (C_i \times P_i) + \gamma X_{ijkt} + \varepsilon_{ijkt} \quad (2)$$

where:

- $W_{ijkt}$ represents real wages
- $P_i$ denotes country-specific AI preparedness index
- Other variables are defined as in equation (1)

We further analyze polarization patterns using a quadratic specification in skill rank:

$$\Delta E_{jk} = \alpha + \beta\ R_j + \beta\ R_j^2 + \beta\ A_k + \beta\ (R_j \times A_k) + \varepsilon_{jk} \quad (3)$$

where:

- $\Delta E_{jk}$ represents change in employment share
- $R_j$ is the skill rank of occupation j
- Other variables are defined as above

This approach extends the traditional Mincerian framework (Mincer, 1974) to incorporate both automation risk and AI preparedness, following recent methodological innovations by Acemoglu et al. (2022).

### 3.4 Identification Strategy

Our identification strategy relies on several key assumptions that we validate through multiple robustness checks. Following the approach of Angrist and Pischke (2009), we address potential threats to identification through a combination of fixed effects, instrumental variables, and falsification tests.

The first key assumption is the comparability of skill levels across countries. We validate this using the ISCO-08 framework and conduct sensitivity analyses following the methodology of Freeman and Oostendorp (2020). This involves testing the stability of our skill classification coefficient $\beta_{S_j}$ across different institutional contexts. The second assumption concerns the validity of automation risk measures $A_k$, which we test using the approach developed by Frey and Osborne (2017) and refined by Webb (2020).

For each occupation j, we construct an automation risk score $A_j$ following:

$$A_j = \sum_{m=1}^{M} w_m T_{jm} \quad (4)$$

where:

- $T_{jm}$ represents the score for task characteristic m in occupation j
- $w_m$ represents the weight assigned to task characteristic m
- M is the total number of task characteristics considered

For causal interpretation of our estimates, we employ an instrumental variables approach using historical technology adoption patterns as instruments for current automation risk, following the methodology of Acemoglu and Restrepo (2019). This helps address potential endogeneity concerns arising from simultaneous determination of employment patterns and automation adoption.

### 3.5 Robustness Checks

We conduct an extensive set of robustness checks to validate our main findings. Following Goldberg and Pavcnik (2007), we test the sensitivity of our results to alternative specifications of skill categories and automation risk measures. For each specification s, we estimate:

$$\ln(E_{ijkt}^s) = \alpha^s + \beta^s C_i + \beta^s S_j^s + \beta^s A_k^s + \beta^s (C_i \times S_j^s) + \varepsilon_{ijkt}^s \quad (5)$$

where superscript s denotes the alternative specification being tested. We also employ a series of placebo tests following the methodology of Autor et al. (2020) to verify that our results are not driven by spurious correlations. For placebo tests p, we estimate:

$$\ln(E_{ijkt}) = \alpha + \beta C_i + \beta S_j + \beta A_k^p + \gamma X_{ijkt} + \varepsilon_{ijkt} \quad (6)$$

where $A_k^p$ represents randomly assigned automation risk categories.

To address potential measurement error in automation risk assessment, we follow the approach of Webb (2020) in constructing alternative risk measures based on different combinations of task characteristics. We also test the robustness of our results to alternative classifications of AI preparedness following the methodology of Korinek and Stiglitz (2021).

### 3.6 Limitations

Several limitations of our analysis warrant discussion. First, following Banerjee and Duflo (2011), we acknowledge that our data may not fully capture informal sector dynamics, which are particularly important in developing economies. Second, the two-country comparison, while informative, may not capture the full range of development experiences, as noted by Goldberg and Pavcnik (2007).

Additionally, the rapid evolution of AI capabilities means that our automation risk measures may need continuous updating, a challenge recognized by Acemoglu and Restrepo (2022). We address these limitations through extensive robustness checks and by being transparent about the bounds of our analysis.

## 4. Results

### 4.1 Employment Distribution Patterns

Our analysis reveals substantial differences in labor market structure between India and the United States, particularly when accounting for automation risk. As shown in Table 1 (see Appendix A), following the methodological approach of Autor et al. (2020), we find that India shows significantly higher concentration in low-skill employment (57.43%) compared to the US (24.42%, $p < 0.01$). This pattern aligns with previous findings by Vashisht and Dubey (2019) on developing economy labor markets but suggests even stronger concentration than earlier studies indicated.

The distribution of employment across skill categories shows persistent differences that remain stable over our study period (2018-2023). As detailed in Table 2, Panel A, high-skill employment shares display marked divergence between the two countries (8.37% in India vs. 33.45% in the US, $p < 0.01$), consistent with patterns documented by Das and Kar (2018). Importantly, we find that automation risk scores are significantly higher in India (mean 29.84) compared to the US (mean 25.63, $p < 0.01$), suggesting greater vulnerability to technological displacement.

Regression analysis of employment distributions, following the specification in Equation 1, reveals strong country effects in employment patterns. As reported in Table 3, the coefficient on the interaction between country and automation risk ($\beta = -0.183$, $p < 0.01$) indicates that India's employment is more concentrated in automation-vulnerable occupations. This finding extends previous work by Frey and Osborne (2017) by demonstrating how automation risk varies systematically with development stage.

### 4.2 Wage Premium Analysis

Our wage premium analysis, based on Equation 2, reveals significant differences in returns to skill across development stages. As presented in Table 4, Panel A, we find substantial wage premiums for high-skill workers ($\beta = 0.892$, $p < 0.01$) and medium-skill workers ($\beta = 0.564$, $p < 0.01$) relative to low-skill workers, with these premiums being significantly larger in India. This pattern aligns with findings by Goldberg and Pavcnik (2007) on skill premiums in developing economies but suggests stronger effects when accounting for automation risk.

The interaction between automation risk and wages shows a significant negative relationship ($\beta = -0.145$, $p < 0.01$), while AI preparedness demonstrates a positive association with wage levels ($\beta = 0.234$, $p < 0.01$). Following the analytical framework of Acemoglu and Restrepo (2019), we interpret these results as evidence that technological readiness significantly moderates the relationship between automation risk and labor market outcomes.

Annual wage growth patterns, summarized in Table 5, reveal stronger skill premiums in India, particularly for high-skill workers (6.8% vs. 4.2% in the US, $p < 0.01$). However, as shown in Panel B of the same table, occupations with high automation risk show significantly slower wage growth in both countries, with a more pronounced effect in India. This finding extends work by Webb (2020) on differential impacts of automation across skill levels.

### 4.3 Labor Market Polarization Analysis

Our analysis of employment share changes reveals distinct patterns of polarization between developing and developed economies. While both countries show declining medium-skill employment shares (India: -0.036%, US: -0.052% annually, $p < 0.01$), the U.S. exhibits growth in both high-skill and low-skill employment, consistent with classic polarization patterns documented by Autor and Dorn (2013). In contrast, India shows significant growth only in high-skill employment (0.042% annually, $p < 0.01$), suggesting what we term "incomplete polarization."

The relationship between skill levels and employment changes follows a U-shaped pattern in the US, consistent with findings by Goos et al. (2009). However, India's pattern is more linear, with monotonically increasing employment shares as skill levels rise. This difference suggests that developing economies may experience distinct patterns of technological disruption, supporting theoretical predictions by Korinek and Stiglitz (2021).

### 4.4 Vulnerability Analysis

Our novel vulnerability analysis reveals systematic differences in technological risk across development stages. As documented in Table 6, India faces significantly higher vulnerability scores across all skill levels, with the gap being particularly pronounced for low-skill workers (0.823 vs. 0.312 in the US, $p < 0.01$). This pattern suggests what we term a "double vulnerability" - both from concentration in lower-skill occupations and from lower AI preparedness scores.

Following the methodology of Acemoglu et al. (2022), we find that the interaction between development stage and automation risk produces compound effects. The coefficient on the interaction between country and AI preparedness ($\beta = -0.167$, $p < 0.01$) indicates that lower technological readiness amplifies the negative effects of automation risk in developing economies.

### 4.5 Robustness and Sensitivity Analysis

Our main findings prove robust to multiple alternative specifications and sensitivity checks. As reported in Tables 7 and 8 (see Appendix B), following Goldberg and Pavcnik (2007), we test our results using different skill classification schemes and find consistent patterns across specifications. The employment distribution effects remain significant when using alternative measures of automation risk ($\beta$ ranging from -0.148 to -0.162, all $p < 0.01$).

Instrumental variable estimates using historical technology adoption patterns as instruments for current automation risk, following Acemoglu and Restrepo (2019), confirm our main findings. The IV estimates are slightly smaller in magnitude but remain statistically significant, suggesting that our baseline estimates may provide an upper bound for the true effects.

Our results also remain stable when excluding the pandemic period (2020-2021), addressing concerns about temporary disruptions affecting our findings. The core coefficients change by less than 10% when excluding these years, suggesting that our results capture structural patterns rather than cyclical effects.

## 5. Discussion and Policy Implications

### 5.1 Interpretation of Key Findings

Our analysis reveals fundamental differences in labor market structures between developing and developed economies that extend beyond traditional development theory. The stark disparity in employment distribution between India and the United States—with India showing 57.43% low-skill employment compared to the US's 24.42%—aligns with previous findings on developing economy labor markets (Banerjee and Duflo, 2011; Fields, 2011). However, our results suggest that this concentration is more persistent than earlier studies indicated, particularly when accounting for automation vulnerability. This pattern reflects what Acemoglu and Restrepo (2019) describe as differential technology adoption paths, where developing economies face distinct constraints in transitioning toward skill-intensive production.

The wage premium findings provide new insights into skill-biased technological change in developing economies. The higher returns to skill in India (6.8% annual growth for high-skill wages versus 4.2% in the US) extend beyond the patterns documented by Katz and Autor (1999) on early-stage development. Our results suggest that the interaction between skill premiums and technological factors appears stronger than previously recognized in the literature (Freeman and Oostendorp, 2020). This interaction may create self-reinforcing patterns of inequality, as suggested by recent theoretical work (Acemoglu et al., 2022).

## 5.2 Technological Disruption and Development

The distribution of automation risk across skill levels reveals important patterns for development policy. The concentration of automation risk in middle-skill occupations (mean risk score 29.74 for Job Zone 2) supports Webb's (2020) argument that AI may affect labor markets differently than previous technological changes. However, the significantly higher vulnerability scores in India across all skill levels (0.823 for low-skill versus 0.312 in the US) suggest what we term a "double vulnerability"—both from occupational structure and lower technological absorption capacity.

The role of AI preparedness in wage determination (coefficient 0.234) highlights what Korinek and Stiglitz (2021) identify as a potential source of divergence between developed and developing economies. Our results suggest that the traditional advantage of technological leapfrogging, documented by Comin and Mestieri (2018), may be less applicable in the AI era due to the importance of complementary institutional capabilities. This finding aligns with recent work by Rodrik (2021) on premature deindustrialization and technological capacity.

## 5.3 Policy Implications

### 5.3.1 Skill Development and Education

Our findings suggest the need for significant modifications to traditional approaches to skill development in developing economies. While previous work emphasized formal education (Psacharopoulos and Patrinos, 2018), our results indicate that effective skill development policies must simultaneously address three key dimensions. First, following Acemoglu and Autor (2021), programs must focus on building AI-complementary skills rather than just traditional technical capabilities. Second, as emphasized by Banerjee et al. (2021), policies must protect vulnerable workers in high-risk occupations through targeted training programs. Third, drawing on insights from Goldin and Katz (2020), institutional capacity for continuous skill updating becomes crucial in rapidly evolving technological environments.

### 5.3.2 Technology Adoption and Industrial Policy

The differential patterns of automation risk and wage growth we document suggest that developing economies need targeted approaches to technology adoption. Building on Rodrik's (2016) framework for industrial policy, our findings support three key policy directions. First, governments should promote selective automation in sectors where developing economies have comparative advantages, as suggested by Diao et al. (2021). Second, following Lall and Pietrobelli (2019), countries need to develop robust institutional frameworks for technology absorption. Third, as emphasized by Harrison et al. (2020), creating incentives for investments in AI-complementary skills and infrastructure becomes crucial.

### 5.3.3 Labor Market Institutions

Our evidence on wage determination suggests that labor market institutions play a crucial role in mediating technology's impact. Building on work by Botero et al. (2004), we find that effective institutional responses should include three elements. First, as suggested by World Bank (2021), strengthened social protection systems for workers in high-risk occupations become essential. Second, following Autor and Dorn (2022), flexible certification systems for emerging skills help labor markets adapt. Third, drawing on Freeman (2020), enhanced labor market information systems that track technological change enable better policy responses.

## 5.4 Future Trajectories

The interaction between development stage and technological change suggests several possible trajectories for developing economies. Following recent theoretical work by Acemoglu and Restrepo (2022), we identify three potential paths. First, "divergent development" may occur where technological gaps widen due to differential AI absorption capabilities. Second, "selective convergence" might emerge where some sectors achieve rapid catch-up while others lag. Third, "institutional leapfrogging" could enable developing economies to build new institutions better suited to the AI era.

These trajectories align with recent research on development patterns. Card et al. (2022) document similar divergent patterns in skill premium evolution, while Derenoncourt et al. (2021) find evidence for selective convergence in specific sectors. The institutional leapfrogging possibility finds support in work by Acemoglu et al. (2023) on adaptive institutions in developing economies.

## 5.5 Limitations and Future Research

Several important limitations of our analysis suggest directions for future research. First, while our data provide detailed occupation-level information, they may not fully capture informal sector dynamics, which Banerjee and Duflo (2011) identify as crucial in developing economies. Future work could extend our analysis using matched employer-employee data, following approaches pioneered by Card et al. (2018).

Second, our two-country comparison, while informative, may not capture the full range of development experiences. Extensions to middle-income countries could provide additional insights into transition dynamics, building on work by Goldberg and Pavcnik (2007) on emerging market labor markets. This could help identify critical transition points in technological absorption capacity.

Third, the rapid evolution of AI capabilities means that our automation risk measures may need continuous updating. Future research could develop dynamic measures of automation risk that account for technological progress, following methodological approaches suggested by Frey and Osborne (2017) and refined by recent work on AI capabilities (Acemoglu et al., 2023).

## 6. Conclusion

This paper provides novel evidence on how labor market structures vary systematically with economic development in the age of artificial intelligence and automation. Through a detailed comparison of India and the United States, incorporating new data on occupation-specific automation risk and AI preparedness, we document substantial differences in employment distributions, wage determination patterns, and labor market polarization. Our analysis yields four main contributions to the literature.

First, we provide systematic evidence that employment distributions differ significantly between developing and developed economies, with these differences proving more persistent than previous research suggested. While theories of structural transformation (Lewis, 1954; Harris and Todaro, 1970) predict gradual convergence in employment structures, our findings indicate that significant differences persist even at relatively advanced stages of development. This disparity is amplified by differential automation risks, with developing economies showing higher concentration in automation-vulnerable occupations. This suggests that the path dependence in development trajectories may be stronger than previously recognized (Acemoglu and Robinson, 2012).

Second, our analysis of wage premiums reveals that returns to skill vary systematically with both development stages and technological readiness. While previous work documented higher returns to education in developing countries (Psacharopoulos and Patrinos, 2018), our findings suggest that these premiums reflect broader structural features of developing economy labor markets. The interaction between skill scarcity and technological factors appears particularly important, as evidenced by the significant positive relationship between AI preparedness and wage levels. This supports recent theoretical work on labor market institutions in development (Freeman and Oostendorp, 2020) while highlighting new dimensions of wage inequality.

Third, we demonstrate that labor market polarization follows distinct patterns across development stages when accounting for automation potential. While the "hollowing out" of middle-skill jobs is well-documented in advanced economies (Autor et al., 2020), we find that developing economies face different challenges. The incomplete polarization we document in India—characterized by high-skill employment growth without corresponding growth in service-sector low-skill jobs—suggests that theories of technological change and employment need modification when applied to developing country contexts. This finding has particular relevance as countries navigate the transition to AI-intensive production systems.

Fourth, our vulnerability analysis introduces a new framework for understanding the interaction between development stage and technological disruption. The significantly higher vulnerability scores across all skill levels in developing economies suggest what we term a "double vulnerability"—both from occupational structure and lower technological absorption capacity. This extends recent work on automation and labor markets (Frey and Osborne, 2017; Webb, 2020) by highlighting development-specific challenges in technological adaptation.

These findings have important implications for both research and policy. For researchers, they highlight the need to account for both development stages and technological capabilities when analyzing labor market phenomena. Models and theories developed for advanced economies may need significant modification when applied to developing country contexts, particularly in understanding the interaction between skill development, technological adoption, and labor market outcomes (Acemoglu and Restrepo, 2022).

For policymakers, our results suggest that effective interventions need to simultaneously address traditional development challenges and emerging technological disruptions. Following recent policy frameworks (World Bank, 2021), this includes developing new approaches to skill development that emphasize AI-complementary capabilities while protecting vulnerable workers; creating institutional frameworks that can facilitate technological absorption while managing displacement risks; and implementing social protection systems that account for both traditional labor market vulnerabilities and new technological risks.

Looking forward, our findings suggest several promising directions for future research. First, extending this analysis to a broader range of countries could help identify critical transition points in development trajectories, building on work by Goldberg and Pavcnik (2007). Second, developing dynamic measures of automation risk and AI preparedness could improve our understanding of how technological capabilities evolve, following methodological innovations by Webb (2020). Third, exploring the causal mechanisms linking technological absorption and labor market outcomes could inform more effective policy interventions, extending recent work on policy evaluation in developing economies (Duflo et al., 2021).

Understanding how labor markets evolve with development remains crucial for both theory and policy. Our findings suggest that while some labor market challenges are universal, their manifestation and appropriate policy responses vary systematically with development stages and technological capabilities. This implies that effective policy design requires careful attention to context-specific factors rather than wholesale adoption of approaches that proved successful in advanced economies. As countries navigate the transition to AI-intensive production systems, this nuanced understanding of development-specific challenges and opportunities becomes increasingly important for fostering inclusive growth and sustainable development.

# Appendix A: Main Results Tables

## Table 1: Employment Distribution by Skill Level and Country (2018-2023)

[As referenced in Section 4.1]

| Skill Level | India (%) | United States (%) | Difference | p-value |
|---|---|---|---|---|
| Low-Skill | 57.43 | 24.42 | 33.01 | <0.01 |
| Medium-Skill | 34.20 | 42.13 | -7.93 | <0.01 |
| High-Skill | 8.37 | 33.45 | -25.08 | <0.01 |
| Total | 100.00 | 100.00 | - | - |

Note: Percentages represent share of total employment. p-values from two-tailed t-tests of difference in means.

## Table 2: Employment Shares and Growth Analysis

[As referenced in Section 4.1]

**Panel A: Employment Shares and Growth Rates by Skill Level (2018-2023)**

| Skill Level | Employment Share (%) | Annual Growth Rate (%) | Automation Risk Score |
|---|---|---|---|
| **India** | | | |
| High-Skill | 8.37 | +0.42 | 23.45 |
| Medium-Skill | 34.20 | -0.36 | 29.74 |
| Low-Skill | 57.43 | -0.06 | 36.32 |
| **United States** | | | |
| High-Skill | 33.45 | +0.38 | 19.82 |
| Medium-Skill | 42.13 | -0.52 | 25.63 |
| Low-Skill | 24.42 | +0.14 | 31.44 |

**Panel B: Summary Statistics of Key Variables**

| Variable | India Mean (SD) | US Mean (SD) |
|---|---|---|
| AI Preparedness Index | 42.6 (-) | 78.3 (-) |
| Automation Risk Score | 29.84 (8.62) | 25.63 (7.45) |
| Annual Wage Growth | 4.83 (1.82) | 3.27 (0.92) |

**Table 3: Employment Distribution Regression Results**

[As referenced in Section 4.1]

| Variable | Coefficient | Std. Error | p-value |
|---|---|---|---|
| Country (India=1) | -0.183 | 0.042 | <0.01 |
| Skill Level | 0.245 | 0.036 | <0.01 |
| Automation Risk | -0.145 | 0.028 | <0.01 |
| Country × Automation Risk | -0.167 | 0.038 | <0.01 |
| AI Preparedness | 0.234 | 0.045 | <0.01 |
| R-squared | 0.684 | | |
| N | 12,450 | | |

**Table 4: Wage Premium Analysis**

[As referenced in Section 4.2]

**Panel A: Wage Premium by Skill Level**

| Skill Level | Wage Premium (%) | Std. Error | p-value |
|---|---|---|---|
| **India** | | | |
| High-Skill | 89.2 | 0.068 | <0.01 |
| Medium-Skill | 56.4 | 0.054 | <0.01 |
| Low-Skill (base) | - | - | - |
| **United States** | | | |
| High-Skill | 72.3 | 0.056 | <0.01 |
| Medium-Skill | 41.2 | 0.043 | <0.01 |
| Low-Skill (base) | - | - | - |

**Table 5: Annual Wage Growth Analysis (2018-2023)**

[As referenced in Section 4.2]

**Panel A: By Skill Level**

| Category | India (%) | United States (%) | Difference | p-value |
|---|---|---|---|---|
| High-Skill | 6.8 | 4.2 | 2.6 | <0.01 |
| Medium-Skill | 4.5 | 3.1 | 1.4 | <0.01 |
| Low-Skill | 3.2 | 2.5 | 0.7 | <0.01 |

**Panel B: By Automation Risk**

| Category | India (%) | United States (%) | Difference | p-value |
|---|---|---|---|---|
| High Risk | 2.8 | 2.1 | 0.7 | <0.01 |
| Medium Risk | 4.3 | 3.4 | 0.9 | <0.01 |
| Low Risk | 5.9 | 4.8 | 1.1 | <0.01 |

**Table 6: Vulnerability Analysis by Skill Level and Country**

[As referenced in Section 4.4]

| Skill Level | India | United States | Difference | p-value |
|---|---|---|---|---|
| High-Skill | 0.534 | 0.223 | 0.311 | <0.01 |
| Medium-Skill | 0.678 | 0.287 | 0.391 | <0.01 |
| Low-Skill | 0.823 | 0.312 | 0.511 | <0.01 |
| Overall Mean | 0.678 | 0.274 | 0.404 | <0.01 |

Note: Vulnerability scores range from 0 to 1, with higher scores indicating greater vulnerability.

**Appendix B: Robustness Checks and Alternative Specifications**

**Table 7: Robustness Checks - Alternative Skill Classifications**

[As referenced in Section 4.5]

| Specification | Employment Effect | Wage Premium Effect | N |
|---|---|---|---|
| Baseline | -0.183*** | 0.892*** | 12,450 |
| Alternative 1 | -0.176*** | 0.878*** | 12,450 |
| Alternative 2 | -0.162*** | 0.865*** | 12,450 |
| Alternative 3 | -0.148*** | 0.854*** | 12,450 |

Note: *** p<0.01, ** p<0.05, * p<0.1. Standard errors clustered at occupation level.

**Table 8: Robustness Checks - Alternative Automation Risk Measures**

[As referenced in Section 4.5]

| Risk Measure | Employment Effect | Wage Effect | AI Preparedness Effect |
|---|---|---|---|
| Baseline | -0.183*** | -0.145*** | 0.234*** |
| Task-Based | -0.175*** | -0.138*** | 0.228*** |
| Industry-Adjusted | -0.169*** | -0.142*** | 0.231*** |
| Occupation-Specific | -0.178*** | -0.144*** | 0.229*** |

Note: *** p<0.01, ** p<0.05, * p<0.1. All specifications include full set of controls.